\documentclass[twocolumn,letterpaper,english,preprintnumbers,amsmath,amssymb,superscriptaddress,footinbib,prx]{revtex4-2}  
\usepackage{graphicx}

 \usepackage{url}

\graphicspath{{./Images/}}
\usepackage{xcolor}
\usepackage[colorlinks = true, citecolor=blue, linkcolor=black,urlcolor=purple]{hyperref}
\usepackage{enumitem}
    \setlist[itemize]{noitemsep, topsep=0pt}
    \setlist[enumerate]{noitemsep, topsep=0pt}
\usepackage{verbatim}
\usepackage{mathtools,amssymb,amsmath}
\usepackage{array}
\usepackage{bm}
\usepackage{multirow}

\usepackage{algorithm}
\usepackage{algpseudocode}

\usepackage{tikz}
\usepackage{xcolor}
\usetikzlibrary{patterns}
\allowdisplaybreaks

\usepackage{comment}

\usepackage[normalem]{ulem}

\begin{document}

\title{Synaptic Theory of Chunking in Working Memory}

\author{Weishun Zhong}
\email{wszhong@ias.edu}
 \affiliation{School of Natural Sciences, Institute for Advanced Study, Princeton, NJ, 08540, USA}
\color{black}

\author{Mikhail Katkov}
\email{mikhail.katkov@gmail.com}
 \affiliation{School of Natural Sciences, Institute for Advanced Study, Princeton, NJ, 08540, USA}
 \affiliation{Department of Brain Sciences,
Weizmann Institute of Science, Rehovot, 76100 Israel}

\author{Misha Tsodyks}
\email{misha@weizmann.ac.il}
\affiliation{School of Natural Sciences, Institute for Advanced Study, Princeton, NJ, 08540, USA}
\affiliation{Department of Brain Sciences,
Weizmann Institute of Science, Rehovot, 76100 Israel}\

\begin{abstract}

Working memory often appears to exceed its basic span by organizing items into compact representations called chunks. Chunking can be learned over time for familiar inputs; however, it can also arise spontaneously for novel stimuli. Such on-the-fly structuring is crucial for cognition, yet the underlying neural mechanism remains unclear. Here we introduce a synaptic theory of chunking, in which short-term synaptic plasticity enables the formation of chunk representations in working memory. We show that a specialized population of ``chunking neurons'' selectively controls groups of stimulus-responsive neurons, akin to gating.  As a result, the network maintains and retrieves the stimuli in chunks, thereby exceeding the basic capacity. Moreover, we show that our model can dynamically construct hierarchical representations within working memory through hierarchical chunking. A consequence of this proposed mechanism is a new limit on the number of items that can be stored and subsequently retrieved from working memory, depending only on the basic working memory capacity when chunking is not invoked. Predictions from our model were confirmed by analyzing single-unit responses in epileptic patients and memory experiments with verbal material. Our work provides a novel conceptual and analytical framework for understanding how the brain organizes information in real time. 
\end{abstract}

\maketitle

\section*{Introduction}
Working memory (WM) is hypothesized to be a distinct capacity for holding and manipulating multiple pieces of information, which is crucial for human cognitive abilities such as verbal communication, reading comprehension, and abstract reasoning \cite{baddeley1992working,barak2014working,miller2018working}. Paradoxically, however, people typically cannot simultaneously hold more than four items in WM \cite{cowan2010magical}. For example, repeating several words or digits is practically effortless and mistake-free, but for lists of five random words, people begin making mistakes \cite{klein2005comparative,ward2010examining,grenfell2012examining}. How, then, are people able to process much larger streams of inputs, such as long passages of text or movies? One attractive idea is chunking, i.e., organizing several items into higher-level units \cite{gobet2001chunking,thalmann2019does,mathy2012s,simon1974big,zhang1985stm,saint2000immediate}. Sometimes chunks are stored in long-term memory due to previous experience \cite{miller1956magical,tulving1962concurrent}, e.g., familiar expression like ``Oh my God" or ``Easier said than done" can be processed as a coherent unit rather than individual words. These pre-existing chunks could be thought of as having stable memory representations learned and consolidated over time, and could therefore be encoded and processed as a single item. However, conceptually more challenging is the phenomenon of \textit{spontaneous} chunking, where novel combinations of items are grouped into separate units ``on the fly", as when a phone number is divided into chunks of 2-3 digits each, or words in a sentence are combined into units based on their syntactic role, such as ``--a little boy--was dressed -- in a green shirt". Indeed, this sentence is much easier to remember than a random sequence of nine words. Surprisingly, a minor manipulation like introducing slight pauses between presentations of consecutive groups of items is enough to trigger chunking and the corresponding increase in capacity \cite{maybery2002grouping,henson1998short,kalm2012neural,oberauer2018benchmarks}. In this study, we addressed two inter-related questions inspired by the above considerations: how spontaneous chunking might emerge in the brain and what is (if there is one) the limit for the number of items that can be held in WM when spontaneous chunking is activated.

Neuronal mechanisms of WM and the origin of WM capacity are still under debate. While the most accepted theory assumes that WM is carried by persistent activity of item-specific neurons \cite{fuster1973unit,miyashita1988neuronal,goldman1995cellular,Amit_1995}, we propose that a more economic and robust mechanism is to rely on short-term plasticity (STP) in item-specific \textit{synapses} \cite{mongillo2008synaptic} (see \cite{stokes2015activity} for a recent review of activity-silent WM). When several items are loaded into WM, rather than having all of the neurons persistently active, information could be maintained by periodic reactivations of the corresponding clusters in the form of population spikes \cite{mongillo2008synaptic,mi2017synaptic}. After each reactivation of a certain cluster, the recurrent self-connections in this cluster remain facilitated, allowing it to bounce back after a period of silence when other clusters activate. The largest possible number of co-active clusters, i.e., the WM capacity, is determined in this theory by the longest possible time between consecutive reactivations for each cluster, which in turn depends on STP time constants \cite{mi2017synaptic}. In the current contribution, we extend the STP theory of WM by including longer-lasting forms of facilitation, such as synaptic augmentation (SA) \cite{zucker2002short}. In \cite{mongillo2024synaptic}, it was shown that due to its slow build-up, SA level in recurrent self-connections encodes the order of presentation of stimuli in WM. While SA does not significantly change the maximal possible number of co-activating clusters, i.e., the basic WM capacity, it allows the network to selectively switch some of the clusters off for a longer period of time, without fully erasing information about their prior activity from the recurrent self-connections \cite{mongillo2024synaptic}. Here, we will show that SA enables consecutive chunks to be activated one after another by switching on and off specialized chunking clusters that serve as controls, and in this way enhance the \textit{effective} WM capacity. In the next section, we demonstrate this mechanism in a simplified neural network model of WM and show how much WM capacity can be increased by chunking compared to the basic regime.

\section*{Results}

\subsection*{Network model of working memory and chunking.}

\begin{figure*}[t!] 
    \centering
        \includegraphics[width=0.9\textwidth]{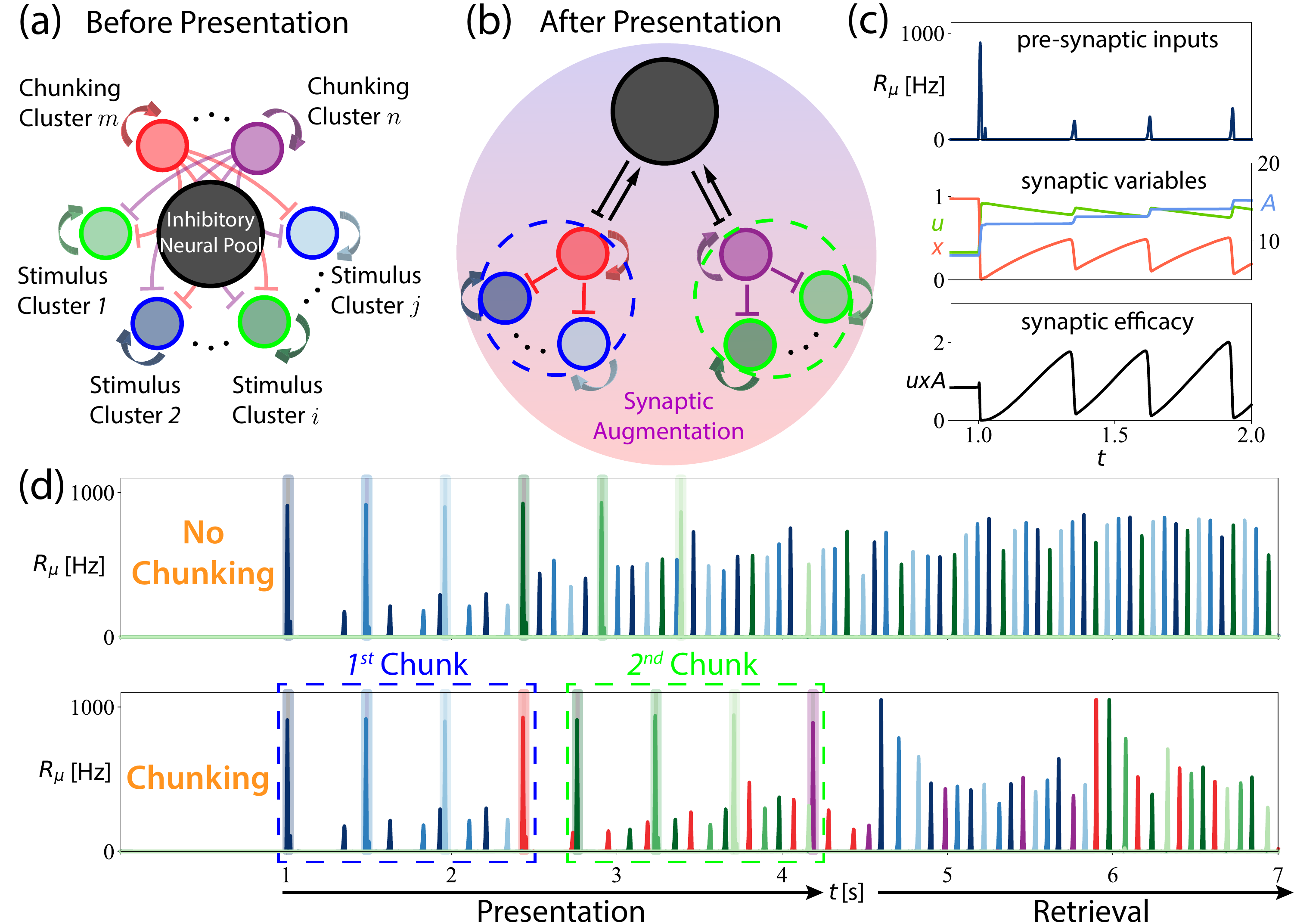}
        \caption{\textbf{Illustration of the hierarchical working memory model.} \\
        \textbf{(a)} Network architecture. Stimulus clusters and chunking clusters both have recurrent self-excitations (thick sharp arrows) and reciprocal connections to the global inhibitory pool (not shown). Chunking clusters have dense but weak connections to the stimulus clusters (thin blunt arrows in the background). \\
        \textbf{(b)} Effective network architecture after presentation. Activities in the network selectively augment connections between stimuli within chunks and the corresponding chunking clusters, effectively forming a hierarchical structure. \\
        \textbf{(c)} Dynamics of the recurrent self-connections. Upon the arrival of pre-synaptic inputs (top panel), the release probability $u$ increases, and the fraction of available neurotransmitters $x$ decreases (left axis of the middle panel). The amplitude of the recurrent strength $A$ gradually increases with each reactivation of the cluster (right axis of the middle panel). As a result, the total synaptic efficacy of the recurrent self-connection $J^{\text{Self}} = uxA$ oscillates (bottom panel). Activity traces are taken from the first stimulus cluster from the top panel of (d) below. \\
        \textbf{(d)} Network simulation. The first three memories are colored in blue, and the other three memories are colored in green. Shades represent external input to the cluster. Top: Memories are loaded at a uniform speed; chunking clusters are not activated. Only four out of six memories remain active in the WM. Bottom: Slight pauses after chunks activate the chunking clusters, which inhibit the stimulus clusters presented before the pause. All memories are retrieved chunk-by-chunk in the retrieval stage. The full activity trace of the synaptic variables is presented in Fig.~\ref{appfig:fig1_extended}. }

\label{fig:schematics}
\end{figure*}

Following our previous work on the synaptic theory of working memory \cite{mongillo2008synaptic,mi2017synaptic,mongillo2024synaptic}, we consider a recurrent neural network model (RNN) where memory items are represented by specific clusters of excitatory neurons coupled to a global inhibitory neural pool, see Fig.~\ref{fig:schematics}(a) and Methods Sec.~\ref{methods:RNN}. The feedback inhibition is assumed to be strong enough such that at any given moment, only one excitatory cluster can be active. To simplify the model, we neglect the overlaps between the stimulus-specific clusters, such that each cluster $\mu$ can be described by a single activity variable corresponding to the average firing rate of the corresponding neurons at a given moment $R_{\mu}(t)$. Furthermore, we assume that all the recurrent self-connections are \textit{dynamic} \cite{markram1998differential, tsodyks1998neural}, i.e., instantaneous synaptic efficacy depends on the pre-synaptic activity within a certain time window due to a combination of short-term synaptic depression and facilitation: $J^{\text{Self}}(t) = u(t)x(t)A$ \cite{markram1998differential}, where $A$ is the amplitude of the recurrent strength, $u(t)$ is the current value of release probability, and $x(t)$ is the current fraction of the maximal amount of neurotransmitter that is available for release.

When the cluster's activity is high, the release probability in the corresponding recurrent connections ($u$) increases above its baseline level $U$, constituting short-term facilitation, and the fraction of available neurotransmitters ($x$) decreases, representing short-term depression (Fig.~\ref{fig:schematics}(c)). When the cluster activity is low, both $u$ and $x$ variables relax towards their baseline values with time constants $\tau_f$ and $\tau_d$, respectively (Methods Sec.~\ref{methods:RNN}). Such transient changes in the synapses are well observed in experiments and are reported to last on the order of hundreds of milliseconds to seconds \cite{zucker2002short,markram1998differential,hempel2000multiple,wang2006heterogeneity}. 

The RNN detailed in Methods Sec.~\ref{methods:RNN} exhibits different dynamical regimes, depending on the STP parameters and external background input. In particular, as shown in \cite{mongillo2008synaptic,mi2017synaptic}, at high background input level, there exists a persistent activity regime where clusters have sustained elevated firing rates corresponding to loaded memory items. As the background input is lowered, there exists a low-activity regime with cyclic behavior where items that were loaded into the network via external stimuli are maintained in WM in the form of sequential brief reactivations called population spikes \cite{tsodyks2000synchrony}. As the number of loaded memories increases, the network eventually fails to maintain some of them, i.e., there is a maximal number of items that can be maintained in the WM, $C$, which depends on the synaptic-level parameters of the RNN \cite{mi2017synaptic}.

In addition to short-term facilitation and depression, experiments observed longer-scale forms of synaptic facilitation in cortical synapses, called synaptic augmentation (SA), characterized by slow, compared to STP, build-up with activity and decay of tens of seconds \cite{hempel2000multiple,wang2006heterogeneity,fisher1997multiple,thomson2000facilitation,fioravante2011short}. We introduce SA as a small transient change in synaptic strength $A$ that is strengthened from its baseline value due to cluster activity, similar to $u$, but with a much longer time constant $\tau_A \gg \tau_f$ \cite{mongillo2024synaptic} (see Fig.~\ref{fig:schematics}(c)).

The main modification of the current model compared to our earlier work is the introduction of distinct excitatory/inhibitory ``chunking" clusters which serve to control the stimulus clusters. Both stimulus clusters and chunking clusters have recurrent excitatory self-connections. Each time the system receives a chunking cue (e.g., when there is a temporal pause in stimulus presentations), one of the chunking clusters is activated and quickly suppresses the currently active stimulus clusters, effectively grouping them into a chunk (Fig.~\ref{fig:schematics}(b)). Subsequent stimulus clusters are then free to be loaded into the network until the next chunking cue is received and another chunking cluster is activated. At the end of the presentation, only chunking clusters reactivate cyclically while all the stimulus clusters are inhibited (Fig.~\ref{fig:schematics}(d)).

\begin{figure*}[t!]
    \centering
        \includegraphics[width=1.05\textwidth]{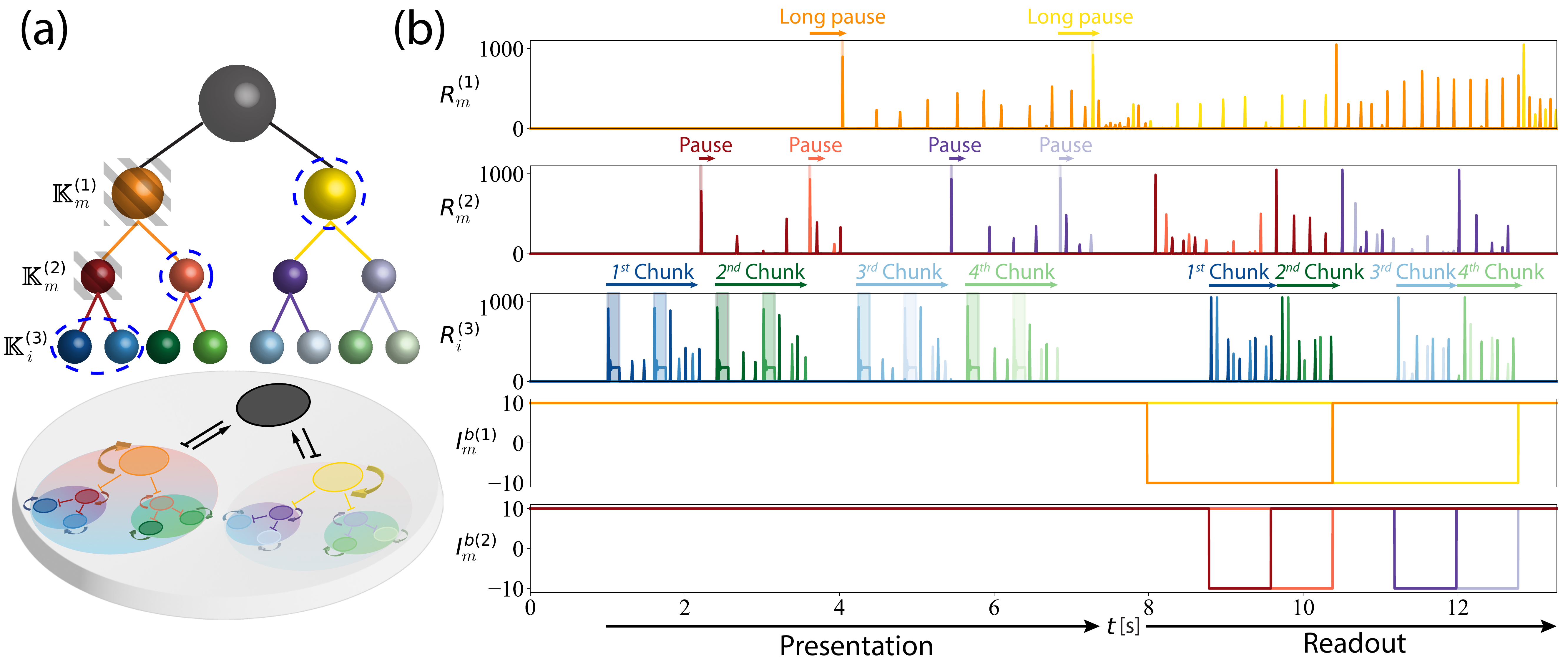}
        \caption{\textbf{Memory retrieval from a hierarchical structure.} \\
        \textbf{(a)} Top: Schematic of an emergent hierarchy of three levels. The top node (black) denotes the global inhibitory neural pool. The first two levels represent chunking clusters, and the lowest level represents stimulus clusters. Grey stripes denote the clusters that need to be suppressed to retrieve the $1^{st}$ chunk. Blue dashed circles represent clusters that are active during the retrieval of the $1^{st}$ chunk during the retrieval stage. Bottom: Architecture of the underlying recurrent neural network. \\
        \textbf{(b)} Simulation of the network in (a). $R^{(k)}$: activity trace of firing rates, color-coded to match the corresponding clusters in (a). The time-course of the traces is labeled as chunks (stimulus clusters), pauses (chunking clusters), and long pauses (meta-chunking clusters). $I^{b(k)}$: activity traces of background input currents. Decreasing the background input to a cluster at level $k$ suppresses its reactivation and removes the inhibition on its children clusters at level $k-1$. }
\label{fig:hierarchy}
\end{figure*}
\subsection*{Chunking increases working memory capacity.} 

The main idea of the proposed chunking mechanism is that the chunking clusters can selectively activate and suppress the stimulus clusters, so that at no point in time do more than a small number of stimulus clusters reactivate as population spikes, thus not exceeding the basic WM capacity. Due to synaptic augmentation, stimulus clusters that are currently suppressed by the chunking clusters still have stronger recurrent self-connections than the ones that were not active at a given trial as long as augmentation has not disappeared. Therefore, the network can \textit{retrieve} temporarily suppressed items by sequentially  switching off the chunking clusters, releasing the suppressed stimulus clusters within the corresponding chunk from inhibition.

To demonstrate the chunking mechanism, we simulate a network of 16 clusters (both stimulus and chunking), 6 of which are activated consecutively with transient external input (presentation stage, the shades in Fig.~\ref{fig:schematics}(d)). We first consider continuous presentation of 6 inputs to the stimulus clusters with no chunking activated. At the end of the presentation, 4 of the corresponding clusters remain active in the form of periodic population spikes while two other clusters drop out of WM, corresponding to a WM capacity of 4 for the chosen values of parameters, similar to \cite{mi2017synaptic} (the top panel of Fig.~\ref{fig:schematics}(d)). Now consider presenting the same six memory items, but with a slightly longer interval between the presentation of the $3^{rd}$ and $4^{th}$ items, during which a chunking cluster is activated (shown in red in the bottom panel of Fig.~\ref{fig:schematics}(d)). We assume that the chunking cluster quickly inhibits the three stimulus clusters that were presented before it (the three blue colors) and remains the only cluster active until the items of the next chunk are presented to the network (shown in green). A second chunking cluster is then activated, shown in purple. This way, the network effectively binds the stimulus clusters in each chunk to their corresponding chunking cluster (Fig.~\ref{fig:schematics}(b)). Such group-specific binding is akin to gating~\cite{soni2025adaptive}, where the activity of each chunking cluster gates the entire chunk of stimulus clusters via inhibition.

We assume that the fast inhibition between chunking clusters and corresponding stimulus clusters happens through strengthening the existing dense but weak inhibitory synapses between them (Fig.~\ref{fig:schematics}(a)). After all stimuli are presented, the network maintains reactivations of two chunking clusters while the synaptic variables of the stimulus clusters slowly decay to their baseline values. However, if the chunking cluster is suppressed within the augmentation time-window $\tau_A$, the items that were inhibited by it will bounce back (Fig.~\ref{fig:schematics}(d) bottom panel, the blue colors in the retrieval stage). At this point in time, four clusters are active: the second chunking cluster and three stimulus clusters from the first chunk, with all items from the first chunk being successfully retrieved. When the second chunk is to be retrieved, the first chunking cluster is again activated by control input while the second chunking cluster is suppressed, allowing the stimulus clusters from the second chunk to activate. This chunking scenario allows the retrieval of all six memory items while at any given moment in time, the network maintains no more than four active clusters, not exceeding basic WM capacity. In this way, chunking increases \textit{effective} working memory capacity by reducing the concurrent load on working memory, at the expense of activating higher-level representations (chunking clusters). 

Above, we chose to illustrate the chunking mechanism in the periodic activity regime because the mechanistic effects of chunking clusters are most apparent with regular firing traces. Nevertheless, our proposed chunking mechanism applies to both the persistent-activity and periodic-activity regimes, with chunking clusters serving the same function in each. Note that, although we model the chunking cues here as slight pauses between presentations, in general chunking can be triggered by other cues, such as tonic variations and semantic meanings. The idea that chunking reduces the load on working memory was first introduced in the psychology literature \cite{miller1956magical,cowan2001magical,thalmann2019does}. Subsequently, neuroimaging studies observed that chunking reduces neural activity in upstream brain regions that process raw stimuli but increases activity in downstream regions associated with higher-level representations \cite{pesenti2001mental,guida2012chunks,kalm2012neural}, which is consistent with our proposed mechanism.

\subsection*{Hierarchical chunking predicts a new capacity.}

\begin{figure*}[t!]
    \centering
        \includegraphics[width=0.9\textwidth]{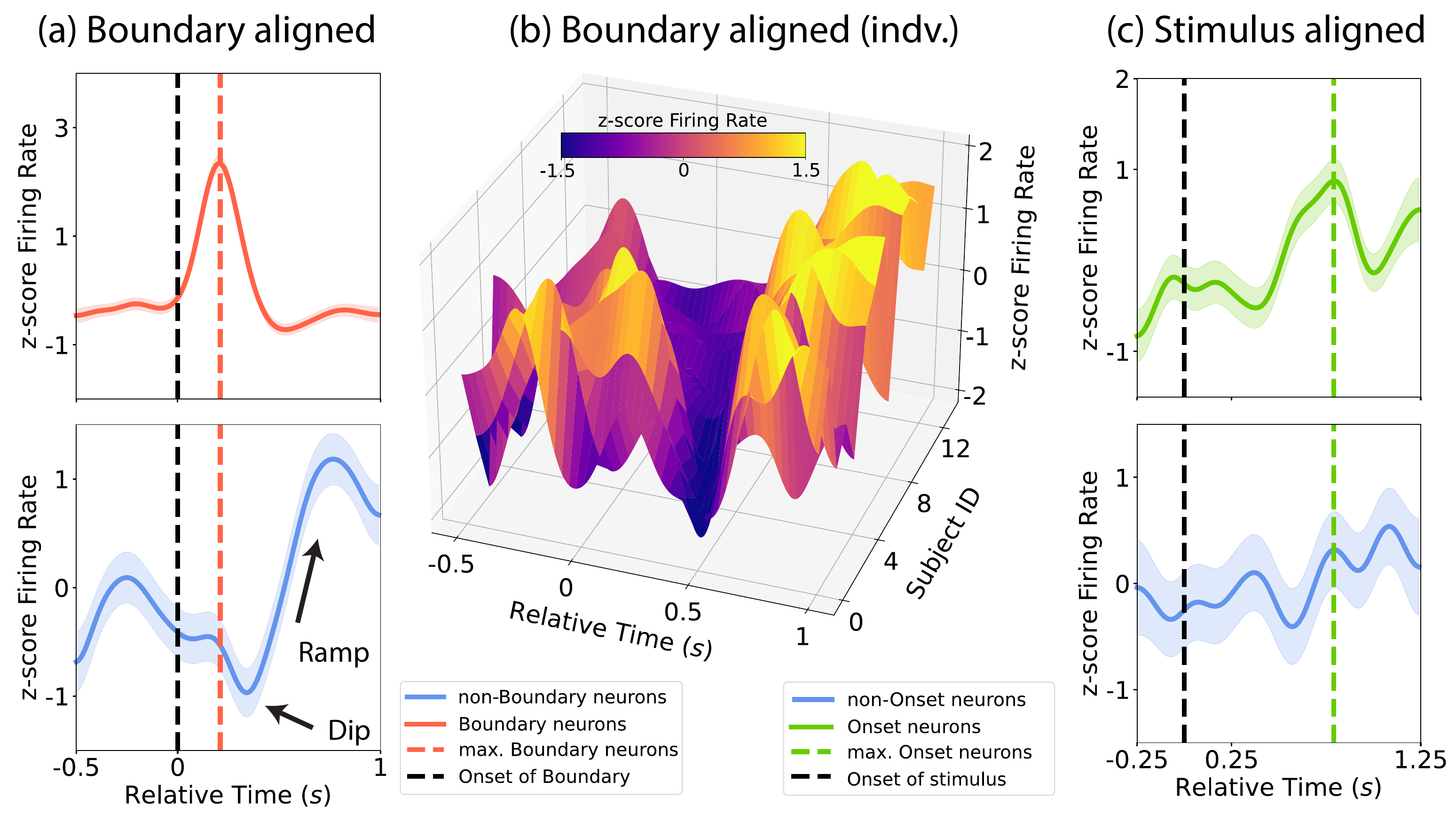}
        \caption{\textbf{Cognitive boundary neurons in the medial temporal lobe.} \\
        \textbf{(a)} Average firing rate from single-neuron recording data in \cite{zheng2022neurons}. The mean z-score firing rates are plotted in solid lines, with one standard deviation included as the shades. Firing rates are averaged over all subjects and trials, and the relative time zero is chosen to be the location of the movie cut. Two qualitative features in the firing rates of the non-boundary neurons: a dip followed by a ramp, are predicted by the hierarchical working memory model. Top: Boundary neurons. Bottom: Non-boundary neurons. \\
        \textbf{(b)} Average firing rates of non-boundary neurons over all trials for individual subjects. Subjects are sorted based on the location of the dip. A trend similar to panel~(a) is observed for each subject. For individual 2D plots, see Fig.~\ref{appfig:fig3_extended}. \\
        \textbf{(c)} Average firing rates of neurons aligned to the onset of the movie (relative time zero). After the peak in onset-specific neurons, the non-onset-specific neurons do not exhibit the dip-then-ramp pattern seen in panel~(a). Top: onset-specific neurons. Bottom: Non-onset specific neurons.} 
\label{fig:boundary_neurons}
\end{figure*}
Our model assumes that several stimulus clusters are grouped into chunks by chunking clusters. A natural question then arises: Can chunks also form meta-chunks? If so, is there a limit to how many levels of such hierarchical representations in working memory can be formed? Here we argue that the answers to both questions are affirmative and moreover, one can derive a surprisingly simple formula for the largest possible number of items in WM (Methods Sec.~\ref{methods:magic}):
\begin{equation}
    \label{main:magic}
    M^* = 2^{C-1},
\end{equation}
where $C$ is the basic WM capacity in the absence of chunking. As mentioned above, $C$ corresponds to the number of active clusters that can be maintained in the RNN model (the top panel of Fig.~\ref{fig:schematics}(d) illustrates the case of $C=4$), and it depends on all the synaptic-level parameters (Methods Sec.~\ref{methods:RNN})\cite{mi2017synaptic,mongillo2024synaptic}.

Eq.~\eqref{main:magic} is a direct consequence of the limited amount of activity that the working memory network can sustain (Methods Sec.~\ref{methods:magic}) and does not depend on specific STP mechanisms. Therefore, we expect $M^*$ to hold in working memory models with similar architecture but possibly different microscopic implementation from Methods Sec.~\ref{methods:RNN}. Eq.~\eqref{main:magic} defines a new capacity for working memory that accounts for hierarchical chunking. Thus, we refer to $M^*$ as the new magic number, in the original spirit of Miller \cite{miller1956magical}.

Below we illustrate how the limited number of $C$ clusters in the network constrains the total number of memory items that can be maintained and retrieved in WM. Let us consider the example corresponding to $C=4$, with a capacity of $M^* = 2^{4-1} = 8$. In this case, the optimal chunking structure is a binary tree with three levels (Fig.~\ref{fig:hierarchy}(a)).

Eight memories are loaded as four chunks of two into the working memory network (the third panel in Fig.~\ref{fig:hierarchy}(b)): a slight pause in-between items of different colors (such as the $2^{nd}$ and $3^{rd}$ items) serves as the chunking cue to activate the chunking clusters (the second panel in Fig.~\ref{fig:hierarchy}(b)), which binds item clusters in pairs of two, similar to the chunks in the bottom panel of Fig.~\ref{fig:schematics}(d). However, here we introduce a slightly longer pause in-between the $4^{th}$ and $5^{th}$ items, during which a chunking cluster binding items 3 and 4 into a chunk is first activated, which is quickly followed by the activation of another chunking cluster to group the first two chunks into a meta-chunk (the first panel in Fig.~\ref{fig:hierarchy}(b)). In this way, after the presentation of the eight items, we have two meta-chunks, giving rise to a tree-like hierarchical structure of three levels (Fig.~\ref{fig:hierarchy}(a)).

\begin{figure*}[t!]
    \centering
        \includegraphics[width=0.7\textwidth]{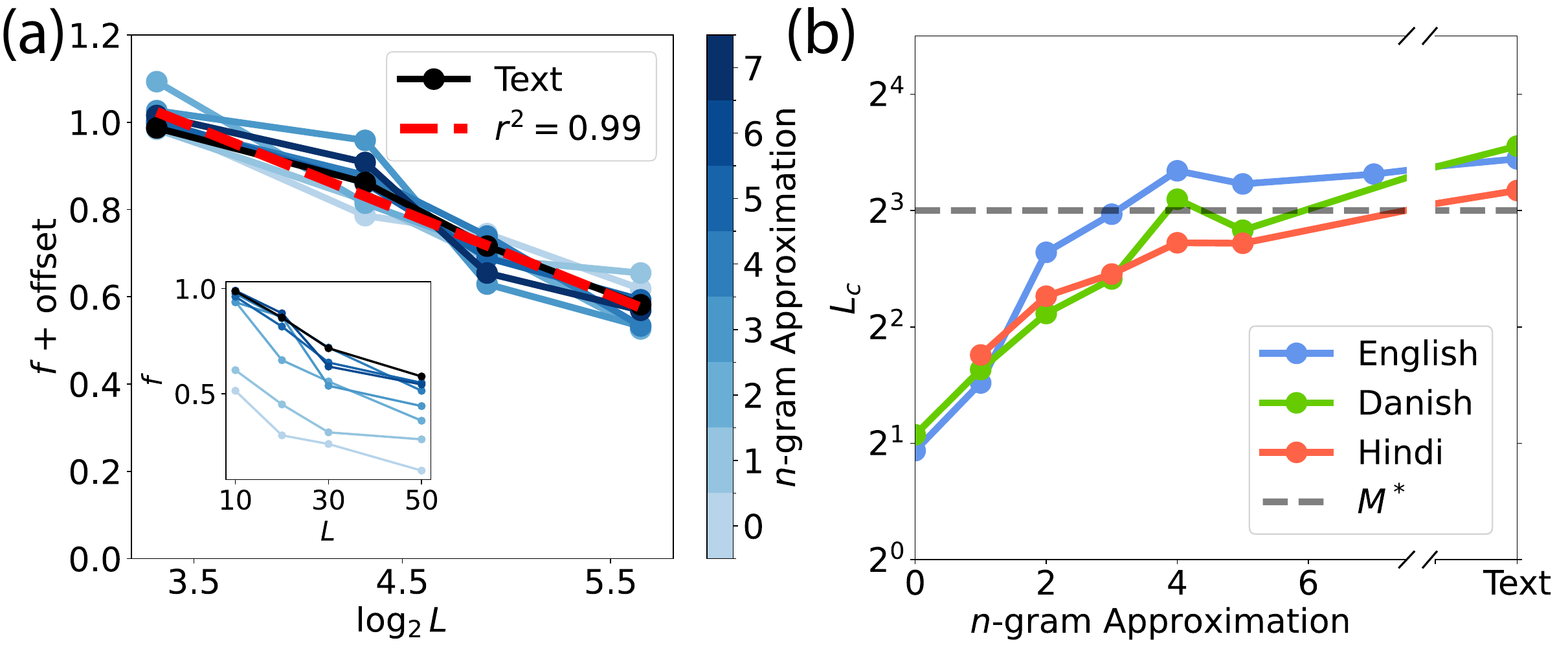}
        \caption{\textbf{The new magic number bounds perfect-recall performance on verbal memory.} \\
        \textbf{(a)} Fraction of recalled words as a function of the length of the presented text. Different shades of blue correspond to different $n$-gram approximations. Black color represents natural text. Inset: Original data as presented in \cite{miller1950verbal}. Main: Different $n$-gram approximation curves become straight lines in a semi-log plot and can be collapsed into a single universal curve (red dashed line) by adjusting the offsets on the individual intercepts. \\
        \textbf{(b)} Critical length of perfect recall as a function of $n$-gram approximations. The location of the critical length $L_c$ is determined by extrapolating the individual $n$-gram approximation curves to where $f(L_c)=1$ using the universal slope. Different colored lines represent experiments in different languages. The grey dashed line corresponds to $M^*=2^{C-1}$ for $C = 4$. }
\label{fig:Miller}
\end{figure*}

To differentiate clusters at different levels of the hierarchy, we denote the $i^{th}$ stimulus clusters at the $k=3$ level as $\mathbb{K}^{(k)}_{i}$ and its activity as $R^{(k)}_i$, where $i$ is the order within the level during presentation. Similarly, the $m^{th}$ chunking clusters at the $k=1,2$ levels is denoted as $\mathbb{K}^{(k)}_{m}$ and its activity as $R^{(k)}_m$. The process of meta-chunking introduced in the previous paragraph can be described more precisely as follows. Immediately after presenting $\mathbb{K}^{(3)}_{4}$ (item 4), chunking cluster $\mathbb{K}^{(2)}_{1}$ is active and suppresses $\mathbb{K}^{(3)}_{1}$ and $\mathbb{K}^{(3)}_{2}$ (items 1 and 2, first two blue colors). Clusters $\mathbb{K}^{(3)}_{3}$ and $\mathbb{K}^{(3)}_{4}$ (items 3 and 4, first two green colors) are also active. Once the chunking signal is received, cluster $\mathbb{K}^{(2)}_{2}$ is activated and suppresses clusters $\mathbb{K}^{(3)}_{3}$ and $\mathbb{K}^{(3)}_{4}$. Furthermore, when this suppression is established, another meta-chunking cluster $\mathbb{K}^{(1)}_{1}$ is activated and suppresses all the $k=2,3$ clusters that were active before it: $\mathbb{K}^{(2)}_{m} (m=1,2)$, $\mathbb{K}^{(3)}_{i} (1\leq i \leq 4)$. When presentation is finished (around $t \sim 7.5$ s in Fig.~\ref{fig:hierarchy}(b)), only the $k=1$ level clusters ($\mathbb{K}^{(1)}_{1}$ in orange, and $\mathbb{K}^{(1)}_{2}$ in yellow) remain active in the working memory.

As the retrieval begins (at $t \gtrsim 8$ s in Fig.~\ref{fig:hierarchy}(b)), $\mathbb{K}^{(1)}_{1}$ is suppressed by a drop in its background input $I^{b(1)}_1$, indicated by grey stripes in Fig.~\ref{fig:hierarchy}(a). Subsequently, the $k=2$ clusters become active, and the network now maintains three clusters: $\{\mathbb{K}^{(2)}_1, \mathbb{K}^{(2)}_2, \mathbb{K}^{(1)}_2\}$. In the next step, $\mathbb{K}^{(2)}_1$ (darker red) is suppressed, which reactivates the $k=3$ stimulus clusters that were inhibited by $\mathbb{K}^{(2)}_1$. Now, the first chunk has been retrieved, and the working memory maintains four clusters: $\{\mathbb{K}^{(3)}_1, \mathbb{K}^{(3)}_2, \mathbb{K}^{(2)}_2, \mathbb{K}^{(1)}_2\}$ (illustrated by blue dashed circles in Fig.~\ref{fig:hierarchy}(a)-(b)). Other chunks can be retrieved in a similar manner by suppressing the corresponding higher-level chunking clusters. Note that even though hierarchical chunking reduces the load on working memory from eight to two, upon unpacking of the chunk, one still needs to maintain all the intermediate chunks that were not unpacked, in this case, $\mathbb{K}^{(2)}_2$ and $\mathbb{K}^{(1)}_2$ (lighter red and yellow, respectively). Similar considerations result in the above expression for $M^*$ in Eq.~\eqref{main:magic} (Methods Sec.~\ref{methods:magic}). In the following sections, we first examine the neuroscience evidence of chunking clusters, then test the prediction of $M^*$ on memory experiments.

\subsection*{Experimental evidence for the existence of chunking clusters.}

Segmentation of sensory stimuli in human memory has been extensively studied in behavioral experiments from the early days of cognitive neuroscience and psychology \cite{miller1956magical,tulving1962concurrent}, but its neural correlates have not been explored until recently \cite{zacks2001human,ben2018hippocampal,kalm2012neural,baldassano2017discovering,zheng2022neurons}. The key assumption in the hierarchical working memory model is the existence of chunking clusters that segment stimuli into chunks. Our model predicts that chunking reduces the load on working memory through inhibition. Upon the firing of the chunking clusters, we expect to see a decrease in the average firing rate of the stimulus clusters. Furthermore, as stimuli continue to be presented after chunking, the average firing rate should gradually increase after the drop. Overall, the hierarchical working memory model predicts two qualitative features in the firing rates of the cluster of neurons that encode stimuli (such as in the bottom panel of Fig.~\ref{fig:schematics}(d)): (1) there should be a ``dip" in the activities of stimulus clusters upon the firing of the chunking clusters; (2) there should be a continuous ``ramping-up" of activities following the dip.

Thanks to advances in single-neuron recording technologies, we can now test our hypothesis using data collected from drug-resistant epilepsy patients \cite{zheng2022neurons}. Consider the experiment reported in \cite{zheng2022neurons}, where subjects are asked to watch a series of movie clips, each consisting of two episodes separated by a ``cut" in the middle of the movie. Such movie cuts serve to induce cognitive boundaries for event segmentation in episodic memory. The authors in \cite{zheng2022neurons} identified a group of neurons in the medial temporal lobe that fire selectively at these boundaries and termed them ``cognitive boundary" neurons. If these neurons segment episodic memories in a manner similar to how chunking clusters segment working memory in our model, then we should also observe a decrease in the firing rates of the stimulus neurons upon the firing of the cognitive boundary neurons. In \cite{zheng2022neurons}, although the boundary neurons can be unambiguously identified by aligning their responses to movie cuts, it is difficult to pinpoint stimulus neurons due to the continuous nature of the visual stimulus. Therefore, we study the putative effect of boundary neurons on the rest of the system by aggregating neurons that are detected but not classified as boundary neurons. We align all the neurons to the movie cuts. Upon averaging over subjects and trials, we find that after the peak in the firing rate of boundary neurons (top panel of Fig.~\ref{fig:boundary_neurons} (a)), about $\sim 130$ ms later, there is a dip in the average activities of the rest of the recorded neurons (bottom panel of Fig.~\ref{fig:boundary_neurons} (a)). Furthermore, there is a continuous ramp-up of activities following the dip. This trend is also evident at the level of individual subjects (Fig.~\ref{fig:boundary_neurons} (b)), and qualitatively agrees with the prediction of our hierarchical working memory model. As a control, within the same recorded population we label the subset of neurons that respond to the onset of the movie clip as ``onset'' neurons. Aligning firing rates to the movie onset, we observe a peak in the onset neurons as reported in \cite{zheng2022neurons}; however, unlike Fig.~\ref{fig:boundary_neurons} (a), the remaining neurons (including boundary neurons) do not exhibit the dip-then-ramp pattern (Fig.~\ref{fig:boundary_neurons} (c)). This indicates that the dip and ramp-up are specific to boundary neurons and suggests an internal network mechanism rather than simple inhibitory feedback or statistical artifacts.

\subsection*{Experimental tests of the new magic number.}

An important prediction of the hierarchical working memory model is the existence of an absolute limit $M^*$, beyond which perfect retrieval is impossible (Eq.~\eqref{main:magic}). One of the earliest studies to quantify this transition is the experiment performed by Miller and Selfridge \cite{miller1950verbal} on the statistical approximation of language. In this experiment, the authors constructed $n$-gram approximations to English, where $n$ refers to coherent occurrences with the previous $n-1$ words. For example, a $1$-gram approximation would consist of words randomly chosen from a corpus. In a $2$-gram approximation, each word would appear coherently with the previous word, but coherence for any sliding window of three words is not required. As $n$ increases, the constructed text gradually approaches natural text. In \cite{miller1950verbal}, subjects were presented with verbal materials constructed from such $n$-gram approximations and asked to recall the words. The fraction of recalled words $f$ decreases with the length $L$ of the material and increases with the degree of approximation $n$ (Fig.~\ref{fig:Miller}(a) inset). Here, we are interested in the critical length $L_c$ beyond which retrieval begins to be imperfect, i.e., $f(L_c)=1$. Since the defining feature of working memory is the ability to perfectly retrieve items that are sustained in the memory, $L_c$ is a measure of working memory capacity. 

In \cite{miller1950verbal}, the fraction of reported words for smallest stimulus length was less than one. To estimate $L_c$, we replotted the data from \cite{miller1950verbal} in a semi-log plot (with $f$ as a function of $\log_2 L$) and observed that all the different $n$-gram curves are well approximated by straight lines. We hence collapsed all the curves into a common line by adjusting the individual intercepts (red dashed line in Fig.~\ref{fig:Miller}(a)). We then used the slope of this line to extrapolate each $n$-gram approximation curve to its critical length $L_c$. We plot $L_c$ as a function of $n$ in Fig.~\ref{fig:Miller}(b). $L_c$ increases with $n$ as expected but starts to plateau around $n=4$, saturating at roughly the predicted value of $8$. Note that $n=0$ corresponds to words randomly chosen from a dictionary, and is dominated by rare words many of which may not be familiar to the subjects. Therefore, the capacity for $n=0$ is expected to be lower than that of common words as in the case for $n=1$.
The same analysis of two replicates of the Miller-Selfridge experiment in Danish and Hindi \cite{maher1988contextual,sharma1977effect} reveals similar trends. As $n$ increases, the verbal material becomes more structured, which allows for the construction of hierarchical representations. Naively, one might expect that the number of perfectly recalled items $L_c$ would continue to increase with $n$, as more structured materials are generally easier to remember. However, we observe that the performance plateaus around $n \sim 4$. This may be due to the fact that longer sentences need to be broken into smaller chunks to be stored in working memory, and there exists an optimal chunk size beyond which storage becomes inefficient and no longer improves memory. This observation qualitatively agrees with our theory in Eq.~\eqref{main:magic}, and the value $n \sim 4$ at which capacity saturates could correspond to the size of a meta-chunk in the optimal hierarchical scheme illustrated above.
Furthermore, our prediction that natural texts are chunked into pairs of two meaningful words resembles the empirical observation of collocations in language, such as adjective-noun, verb-noun, and subject-verb pairs, etc \cite{bartsch2004structural,mel1998collocations,evert2008corpora,mckeown2000collocations}.

Notably, in Fig.~\ref{fig:Miller}(b) for all three languages, $L_c$ saturates within the region predicted by $M^* = 2^{C-1}$, when substituting for $C = 4$ \cite{cowan2010magical}. Therefore, we conclude that the recall performance of verbal materials from working memory agrees with the prediction of our new magic number.

\section*{Discussion}

Chunking is classically believed to be a crucial process for overcoming extremely limited working memory capacity. In the current contribution, we suggest a simple mechanism of chunking in the context of the synaptic theory of working memory. The proposed mechanism relies on the ability of the system to temporarily suppress groups of items without permanently erasing them from WM, which is enabled by the longer-term form of synaptic facilitation, called synaptic augmentation. For chunking to work properly in the model, the system has to utilize separate neuronal clusters, which we call ``chunking clusters" that effectively combine groups of several items each into distinct chunks. Moreover, the activity of chunking clusters has to be controlled in order to allow the suppression and reactivation of subsequent chunks at the right times to avoid saturating working memory capacity at any given moment. In particular, each chunking cluster has to be activated right after all of the corresponding stimuli are presented and later suppressed for them to be retrieved. Our model has no explicit mechanisms for this hypothesized control of chunking clusters; we speculate that it could be triggered by corresponding cues, e.g., chunking clusters could be activated by extra temporal pausing or intonation accentuation, and suppressed by internally generated retrieval signals. While further experimental and theoretical studies are needed to elucidate these suggestions, the existence of specialized chunking neurons has some recent neurophysiological support in electrical recordings in epileptic patients, where neurons responding to cuts in video clips were identified. We analyzed the data collected in these experiments and found that the activity of these and other neurons during clip watching is broadly consistent with our model predictions.

Apart from proposing the biological mechanism of chunking in working memory, we considered the question of whether the hierarchical organization of items in working memory could emerge from the subsequent chunking of chunks. Indeed, we demonstrated that the model allows for such a hierarchical scheme; however, due to working memory capacity, the overall number of items that can be retrieved is still constrained even for the optimal chunking scheme. We derived the universal relation between capacity and the maximal number of retrievable items, which we call a magic number following the classical Miller paper \cite{miller1956magical}. In particular, this relation predicts the new magic number of 8 for a working memory capacity of 4, which is currently accepted as the best estimate of capacity. The chunking scheme achieving this limit corresponds to dividing the inputs into 4 chunks of 2, with two ``meta-chunks'', each consisting of two chunks. We reanalyzed the results of a memory study where subjects were presented with progressively higher-order approximations of meaningful passages for recall, and found that indeed the average maximal number of words that could be fully recalled was close to the predicted value of 8, and that this number saturated for a $4^{th}$ order approximation of meaningful passages, corresponding to the size of a ``meta-chunk" in the optimal chunking scheme predicted by the model. While encouraging, more studies should be performed to elaborate on this issue, in particular to more directly demonstrate the ability of subjects to form chunks of chunks during working memory tasks. \\

Our theory and the proposed neural network mechanism attempt to bridge the microscopic level of neural activities and the macroscopic level of behaviors in the context of hierarchically-structured memories. Our analytical results and data analysis methods offer new perspectives on classical results in cognitive neuroscience and psychology. The proposal of a hierarchical structure in working memory can open many new directions. For instance, long-term memory is usually organized in a hierarchical manner, as reflected in our ability to gradually zoom into increasingly fine details of an event during recall \cite{kurby2008segmentation}. While working memory underlies our ability to construct such hierarchical representations, little is known about how the transient tree-like structure in working memory is related to the hierarchy in long-term memory. Furthermore, one of the hallmarks of fluid intelligence — the ability to compress and summarize information — is also related to re-coding information in a hierarchical manner \cite{chekaf2018compression}. Understanding how our mind is capable of making use of hierarchical structures for complex cognitive functions such as summarization and comprehension remains an important open question.
\\

\acknowledgments{
\textbf{Acknowledgments}. All authors were involved in the conceptualization of the study. WZ conducted the numerical simulations and data analysis. All authors contributed to writing the manuscript. The authors declare no competing interests. All data analyzed in the manuscript are publicly available online. The simulation code will be made available in a GitHub repository upon publication. \\
We want to thank Tankut Can, 
Antonis Georgiou, Máté Lengyel, Gianluigi Mongillo, Sandro Romani, Sebastian Seung, and Jie Zheng for helpful discussions. W.Z. is supported by Eric and Wendy Schmidt Member in Biology Fund and the Simons Foundation at the Institute for Advanced Study. MK is supported in part by a grant from Fran Morris Rosman and Richard Rosman. M.T. is supported by the Simons Foundation, MBZUAI-WIS Joint Program for Artificial Intelligence Research and Foundation Adelis. }

\section*{Methods}

\subsection{RNN model for hierarchical working memory}
\label{methods:RNN}

As illustrated in Fig.~\ref{fig:schematics}(a), the recurrent network that implements WM has 3 functionally distinct types of neuronal populations: stimulus clusters that encode different items (indexed by $i$ below), chunking clusters (indexed by $m$), and a single inhibitory neural pool indexed by $I$.  
WM implementation is based on the previously introduced synaptic theory of working memory \cite{mongillo2008synaptic,mi2017synaptic,mongillo2024synaptic}. All stimulus and chunking clusters exhibit short-term synaptic plasticity in the recurrent self-connections, such that the instantaneous strength of connections for cluster $\mu$ ($\mu = (i,m)$) is given by
\begin{equation}
    J_{\mu \mu}(t) = A_\mu(t) u_\mu(t) x_\mu(t),
\end{equation}
where $A$ is the amplitude of the recurrent strength, $u$ is the probability of release, and $x$ is the fraction of available neurotransmitters; all three factors depend on time via the following dynamical equations reflecting different STP processes:
\begin{align}
    \frac{d u_{\mu}}{dt} &= \frac{U-u_\mu}{\tau_f} + U(1-u_\mu)R_\mu, \label{eq:dynamic_synapses}\\
    \frac{d x_\mu}{dt} &= \frac{1-x_\mu}{\tau_d} - u_\mu x_\mu R_\mu, \label{eq:dynamic_synapses2}\\
    \frac{d A_{\mu}}{dt} &= \frac{A_{\text{min}}-A_\mu}{\tau_A} + \kappa_A(A_{\text{max}}-A_\mu)R_\mu, \label{eq:dynamic_synapses3}   
\end{align}
where $R_{\mu}$ is the activity of cluster $\mu$; $U$ is the baseline value of release probability; $\tau_f$, $\tau_d$ and $\tau_A$ are time constants of synaptic facilitation, depression and augmentation, correspondingly; $A_{\text{min}}$, $A_{\text{max}}$ and $\kappa_A$ are parameters of synaptic augmentation that distinguish this model from earlier versions. Apart from self-connections, each stimulus and chunking cluster is reciprocally connected to the inhibitory pool, and some of the chunking clusters develop quick inhibition on groups of stimulus clusters as explained below. The activity of each cluster is determined as a non-linear gain function of its input, and all inputs satisfy the following standard dynamics:
\begin{align}
    \tau \frac{d h_{\mu}}{dt} &= -h_{\mu} +\sum_{\nu} J_{\mu\nu}(t)R_{\nu} - w^{EI}R_I + I^b_{\mu}(t) + I^e_{\mu}(t), \label{eq:RNN}\\
    \tau \frac{dh_I}{dt} &= -h_I + w^{IE} \sum_{\nu} R_{\nu}, \label{eq:RNN2}
\end{align}
where $R(h) = \alpha \ln (1 + \exp(h/\alpha))$ is a soft threshold-linear gain function mentioned above. $I^b$ stands for external background inputs from other regions of the brain that reflect the general level of activity in the network, and $I^e$ is the external input used to load memory stimuli. $w^{EI}$ and $w^{IE}$ define the strength of feedback inhibition between stimulus and chunk clusters and the global inhibitory cluster. Furthermore, we assume that when a chunking cluster $m$ gets activated by a chunking cue at $t_c$ during the presentation, the weak inhibitory synapses are selectively strengthened between the chunking cluster and the stimulus clusters $i$ in the same chunk presented before it:
\begin{equation}
\label{eq:inhibition}
    J_{im}(t) = J_{\text{inh}}\Theta(t-t_c)\delta_{im}, \qquad i \in \; \text{chunk}\; m.
\end{equation}
See Fig.~\ref{appfig:fig1_extended} for an illustration of the synaptic matrix before and after chunking. For the hierarchical structure in Fig.~\ref{fig:hierarchy}(b), we generalize Eq.~\eqref{eq:inhibition} to higher-level chunking clusters, such that the $k^{th}$ level chunking clusters inhibit all the lower-level clusters presented before them (both chunking and stimulus).

The detailed synaptic mechanism for behavioral time scale plasticity such as Eq.~\eqref{eq:inhibition} is subject to much active research \cite{nicoll2005synaptic, vandael2020short,bittner2017behavioral,milstein2021bidirectional,pang2024non}. Here in the RNN model, we do not attempt to explain its mechanism but rather assume that it takes place via external control. The microscopic implementation of Eq.~\eqref{eq:inhibition} is not crucial to the proposed chunking mechanism, and in Methods Sec.~\ref{methods:additional}, we present additional RNN simulations that adopt a possible implementation of Eq.~\eqref{eq:inhibition} and achieve similar activity traces as in Fig.~\ref{fig:schematics}(d) and Fig.~\ref{fig:hierarchy}(b).

\subsection{The new magic number}
\label{methods:magic}

At any given moment, the network cannot maintain more than $C$ active clusters (Fig.~\ref{fig:schematics}(d) top panel illustrates the case of $C=4$), and we refer to $C$ as the basic working memory capacity. Even though we can potentially encode an arbitrarily deep hierarchical representation, $C$ nevertheless constrains how many stimulus clusters can be retrieved. To understand the consequence of this constraint, we abstract away from the recurrent neural network and consider the effective hierarchical representation entailed by its activity (Fig.~\ref{fig:hierarchy}(a)).

Let us denote the size of the $m^{th}$ chunk at the $k^{th}$ level ($1\leq k \leq K$) as $c_{k_m}$, which is the same as the branching ratio of its parent level. For example, the effective tree-like hierarchical structure in Fig.~\ref{fig:hierarchy}(a) has four chunks of two stimulus clusters at the $k=3$ level. It proves to be instructive to first consider a slightly simplified setting, where at a given level $k$ all the chunk sizes are the same horizontally, $c_{k_m} = c_k$ for all chunks $m$ (e.g., $c_{2_m} = 2$ for all four of the $k=3$ level chunks in Fig.~\ref{fig:hierarchy}(a)). Later, we will relax this assumption and show that the result we derive below still holds.

To retrieve a chunk from the bottom of the hierarchy, i.e., the stimulus clusters that encode actual memories, we need to suppress nodes upstream of the desired chunk. As a result, children of the suppressed node will become reactivated. A series of suppressions from the top to the bottom of the hierarchy requires the working memory to simultaneously maintain $c_K$ stimulus chunks from the bottom level, as well as $c_k - 1$ chunking clusters from each of the $k^{th}$ level above ($1\leq k < K$) that were not suppressed but become active due to the suppression of their parent. However, the total number of clusters that can be maintained must not exceed $C$ (e.g., the total number of clusters enclosed by the blue dashed circles in Fig.~\ref{fig:hierarchy}(a) should not exceed 4),
\begin{equation}
\label{eq:constraint}
    C \geq \sum_{k=1}^{K} (c_k-1) + 1.
\end{equation}
Meanwhile, the total number of stimulus clusters encoded in the hierarchical structure is 
\begin{equation}
\label{eq:total_leaves}
    M = \prod_{k=1}^{K} c_k.
\end{equation}
To achieve maximum capacity, we maximize Eq.~\eqref{eq:total_leaves} subject to the constraint in Eq.~\eqref{eq:constraint}. Using the arithmetic and geometric mean inequality, we arrive at
\begin{equation}
\label{eq:capacity}
    M \leq M_c (K) = \left(1 + \frac{C-1}{K} \right)^K,
\end{equation}
where the equality is saturated when the branching ratio (chunk size) $c_k$ at all levels are equal,
\begin{equation}
    c_k = 1 + (C-1)/K.
\end{equation}
We notice that $M_c(K)$ monotonically increases with $K$. Since the chunk size considered here $c_k$ needs to be an integer, we have the optimal level $K^*$ and optimal branching ratio $c^*$
\begin{equation}
\label{eq:optimal_tree}
    K^* = C-1, \qquad c^*=2.
\end{equation}
Substituting Eq.~\eqref{eq:optimal_tree} into Eq.~\eqref{eq:capacity}, we arrive at the capacity
\begin{equation}
\label{eq:magic_number}
    M^* = 2^{C-1}.
\end{equation}
Next, let us consider relaxing the simplifying assumption of $c_{k_m} = c_k$. Without loss of generality, suppose that at the $k^{th}$ level, $c_{k_m} > c_{k_{m+1}}$. In order to retrieve the $m^{th}$ chunk at this level, the WM needs to at least maintain $c_{k_{m}}$ clusters, which implies that when trying to retrieve the $(m+1)^{th}$ chunk the WM is not saturated because all the levels above the $k^{th}$ are identical for the $m^{th}$ and $(m + 1)^{th}$ chunk. This is sub-optimal since our goal is to maximize $M$. Therefore, $c_{k_{m+1}}$ can be increased to at least as large as $c_{k_m}$. The same logic can be applied recursively to all levels of the hierarchy, which demands that the optimal hierarchical structure for maximum $M$ has $c_{k_m} = c_k$, so we again arrive at $M^*$ in Eq.~\eqref{eq:magic_number}.
\\

\subsection{RNN simulations}
\label{methods:numerics}

Activity traces of all the dynamical variables in Eq.~\eqref{eq:dynamic_synapses}-\eqref{eq:inhibition} are shown in the Fig.~\ref{appfig:fig1_extended}. In particular, the synaptic matrix $J_{\mu\nu}$ before and after chunking in Fig.~\ref{fig:schematics}(d) is shown for comparison. All simulation parameters are reported in Table~\ref{table:params}. All the external inputs $I^e$ used for loading the memories are rectangular functions with support only at the presentation time, and have an amplitude of $750\ \mathrm{Hz}$, and all the background input $I^b$ has amplitude of $|I^b| = 10\ \mathrm{Hz}$. Additionally, the timing of the external control signals are summarized in below.

Fig.~\ref{fig:schematics}(d) top panel: Stimulus starts to load at $t=1\ \mathrm{s}$ for a duration of $0.025\ \mathrm{s}$ with an interval of $0.45\ \mathrm{s}$. Background input $I^b$ has a constant value of $10\ \mathrm{Hz}$.

Fig.~\ref{fig:schematics}(d) bottom panel: Stimulus starts to load at $t=1\ \mathrm{s}$ for a duration of $0.025\ \mathrm{s}$ with an interval of $0.45\ \mathrm{s}$. Chunking clusters are loaded for a duration of $0.025\ \mathrm{s}$ with an interval of $0.3\ \mathrm{s}$. Background input $I^b$ has a constant value of $10\ \mathrm{Hz}$ during the presentation stage and switches between $10\ \mathrm{Hz}$ and $-10\ \mathrm{Hz}$ for a duration of $1.35\ \mathrm{s}$ during the retrieval stage.

Fig.~\ref{fig:hierarchy}(b): The $k=3$ level stimulus clusters start to load at $t=1\ \mathrm{s}$ for a duration of $0.15\ \mathrm{s}$ with an interval of $0.45\ \mathrm{s}$. $k=2,3$ level chunking clusters load for a duration of $0.01\ \mathrm{s}$ with an interval of $0.2\ \mathrm{s}$. Background input $I^b$ has a constant value of $10\ \mathrm{Hz}$ during the presentation stage and switches between $10\ \mathrm{Hz}$ and $-10\ \mathrm{Hz}$ for a duration of $0.8\ \mathrm{s}$ during the retrieval stage.

\begin{table}[h!]
\centering
\renewcommand{\arraystretch}{1.5} 
\begin{tabular}{ |@{\hspace{1em}}c@{\hspace{1em}}|c|c| } 
\hline
Equations & Parameters & Values \\
\hline
\multirow{5}{*}{\makebox[4em][l]{Eq.~\eqref{eq:RNN}-\eqref{eq:RNN2}}} & $\tau$ - Membrane time constant & $8\ \mathrm{ms}$ \\ 
& $w^{EI}$ - Synaptic efficacy $I \to E$ & 1.5 \\ 
& $w^{IE}$ - Synaptic efficacy $E \to I$ & 2.4 \\ 
& $\alpha$ - Gain function parameter & $1.5\ \mathrm{Hz}$ \\ 
& $P$ - Total number of clusters & 16 \\ 
\hline
\multirow{7}{*}{Eq.~\eqref{eq:dynamic_synapses}-\eqref{eq:dynamic_synapses3}} 
& $\tau_f$ - Facilitation time constant & $1.2\ \mathrm{s}$ \\ 
& $\tau_d$ - Depression time constant & $0.45\ \mathrm{s}$ \\ 
& $\tau_A$ - Augmentation time constant & $75\ \mathrm{s}$ \\ 
& $U$ - Facilitation baseline & $0.3$ \\ 
& $A_{\text{min}}$ - Augmentation baseline & 8 \\ 
& $A_{\text{max}}$ - Maximum augmentation & 30 \\ 
& $\kappa_{A}$ - Augmentation rate & 0.03 \\ 
\hline
\multirow{1}{*}{Eq.~\eqref{eq:inhibition}} & $J_{\text{inh}}$ - Inhibition strength & $ 10$ \\ 
\hline
\end{tabular}
\caption{Simulation parameters in Fig.~\ref{fig:schematics}(d) and Fig.~\ref{fig:hierarchy}(b).}
\label{table:params}
\end{table}

\subsection{Additional simulations}
\label{methods:additional}
Eq.~\eqref{eq:inhibition} assumes that chunking clusters can quickly bind with the stimulus clusters from the same chunk. For such binding to be selective, the synapse of the stimulus clusters need to be able to maintain a memory trace of its past activities. In this section, we attempt to provide a possible mechanism. We assume that there is a time-delayed Hebbian-like strengthening on the inhibitory synapses from the chunking clusters to the stimulus clusters. Such strengthening integrates back in time over a window $\tau_s$ ($\tau_f \ll \tau_s \ll \tau_A$) for stimulus clusters that were presented before the activation of the chunking cluster, and strengthens the originally present but weak synapses between them. Given a stimulus cluster $i$ presented within $\tau_s$ before the chunking cluster $m$, the strength of the inhibitory synapses $J_{im}$ between them gets strengthened according to
\begin{align}
\label{eq:binding}
    \frac{d J_{im}}{dt} &= \frac{J_{\text{min}}-J_{im}}{\tau_J} + \kappa_J (J_{\text{max}}-J_{im})\sigma(\bar{R}_i R_{m} - \theta_0), \\
    \bar{R}_{i} &= \int_{t-\tau_{s}}^{t}dt' R_{i}(t'), \nonumber
\end{align}
where $\bar{R}_{i}$ represents a time-delayed synapse that maintains memory traces over a window $\tau_s$, $\sigma(\cdot)$ is a non-linear function chosen to be the same as the gain function for the firing rates, and $\theta_0$ is a threshold that filters out reactivations.

We expect Eq.~\eqref{eq:binding} to work in the regime where the external input to the network during presentation is much stronger than the subsequent reactivations, which is typically the case. Here, the reactivations are filtered out so that they do not contribute to the binding process and form cross-linking between different chunks. Eq.~\eqref{eq:binding} only strengthens the binding between the chunking cluster $m$ and the stimulus cluster $i$ that were presented within the $\tau_s$ time window, but not the stimuli that were presented outside of $\tau_s$ but reactivate during $\tau_s$, which have much weaker amplitudes. As a result, the time-delayed augmentation effectively binds the chunking cluster with the stimulus clusters presented before it within $\tau_s$. Time-delayed synapses were first introduced in the context of memory sequences \cite{sompolinsky1986temporal,kleinfeld1986sequential,hertz2018introduction}, and are found to be related to behavioral time scale synaptic plasticity through dendritic computation \cite{london2005dendritic,bittner2017behavioral,milstein2021bidirectional}. 

As a potential detailed mechanism of Eq.~\eqref{eq:inhibition}, we perform additional RNN simulations with Eq.~\eqref{eq:binding}. We find that Eq.~\eqref{eq:RNN}-\eqref{eq:dynamic_synapses} with Eq.~\eqref{eq:binding}, instead of Eq.~\eqref{eq:inhibition}, is able to approximate the activity traces as in Fig.~\ref{fig:schematics}(d) and Fig.~\ref{fig:hierarchy}(b) (see Fig.~\ref{appfig:memory_trace}). However, it requires fine-tuning between the presentation time and the integration window $\tau_s$, as well as the threshold $\theta_0$. We report the additional parameters used in Eq.~\eqref{eq:binding} below.

Parameters that are independent of the presentation times: $\tau_J = 75\ \mathrm{s}$, $J_{\text{min}}=0$, $J_{\text{max}}=10$, $\kappa_J = 1\ \mathrm{Hz}$. Parameters that depend on the presentation times: Fig.~\ref{appfig:memory_trace}) (a)-(b): $\tau_s = 1.8\ \mathrm{s}$ and $\theta_0 = 7000\ \mathrm{Hz}$. Fig.~\ref{appfig:memory_trace}) (c): threshold $\theta_0$ is chosen to be proportional to the duration of the loading time with the external input: $\theta_0 = 25600\ \mathrm{Hz}$ for $J^{(2)\dashv (3)}$ and $J^{(1)\dashv (3)}$, but for $J^{(1)\dashv (2)}$ is reduced by a factor of five, where we use $J^{(k) \dashv (l)}$ to denote synaptic matrix components that correspond to the inhibition from level $k$ to $l$; Integration window $\tau_s$ is chosen such that adjacent levels are shorter than skip levels: $\tau_s=1.9\ \mathrm{s}$ for adjacent levels ($k=1$ to $k=2$ and $k=2$ to $k=3$) and $\tau_s = 3.1\ \mathrm{s}$ for skip level ($k=1$ to $k=3$).

\subsection{Cognitive boundary neurons}
Two types of boundary neurons are reported in \cite{zheng2022neurons}: neurons that code for soft boundaries (change of camera position after the cut) and neurons that code for hard boundaries (change of movie content after the cut). In the present study, we do not distinguish between the two types of neurons and classify both as boundary neurons. In Fig.~\ref{fig:boundary_neurons}, we pool together the raw firing rates of all the boundary (or non-boundary) neurons from a subject, then perform the z-score averaging across different subjects. We have excluded four subjects out of eighteen in \cite{zheng2022neurons} from our analysis, because in those subjects either no neurons responding to the onset of the movie were detected, or no neurons responding to the onset of the cut were detected. The z-score firing rates of non-boundary neurons from individual subjects are shown in the Fig.~\ref{appfig:fig3_extended}. Data analyzed in Fig.~\ref{fig:boundary_neurons} is downloaded from the DANDI Archive at \url{https://dandiarchive.org/dandiset/000207/0.220216.0323}.
\\

\renewcommand{\thefigure}{S\arabic{figure}}
\setcounter{figure}{0}

\begin{figure*}[t!] 
    \centering
    \includegraphics[width=1\textwidth]{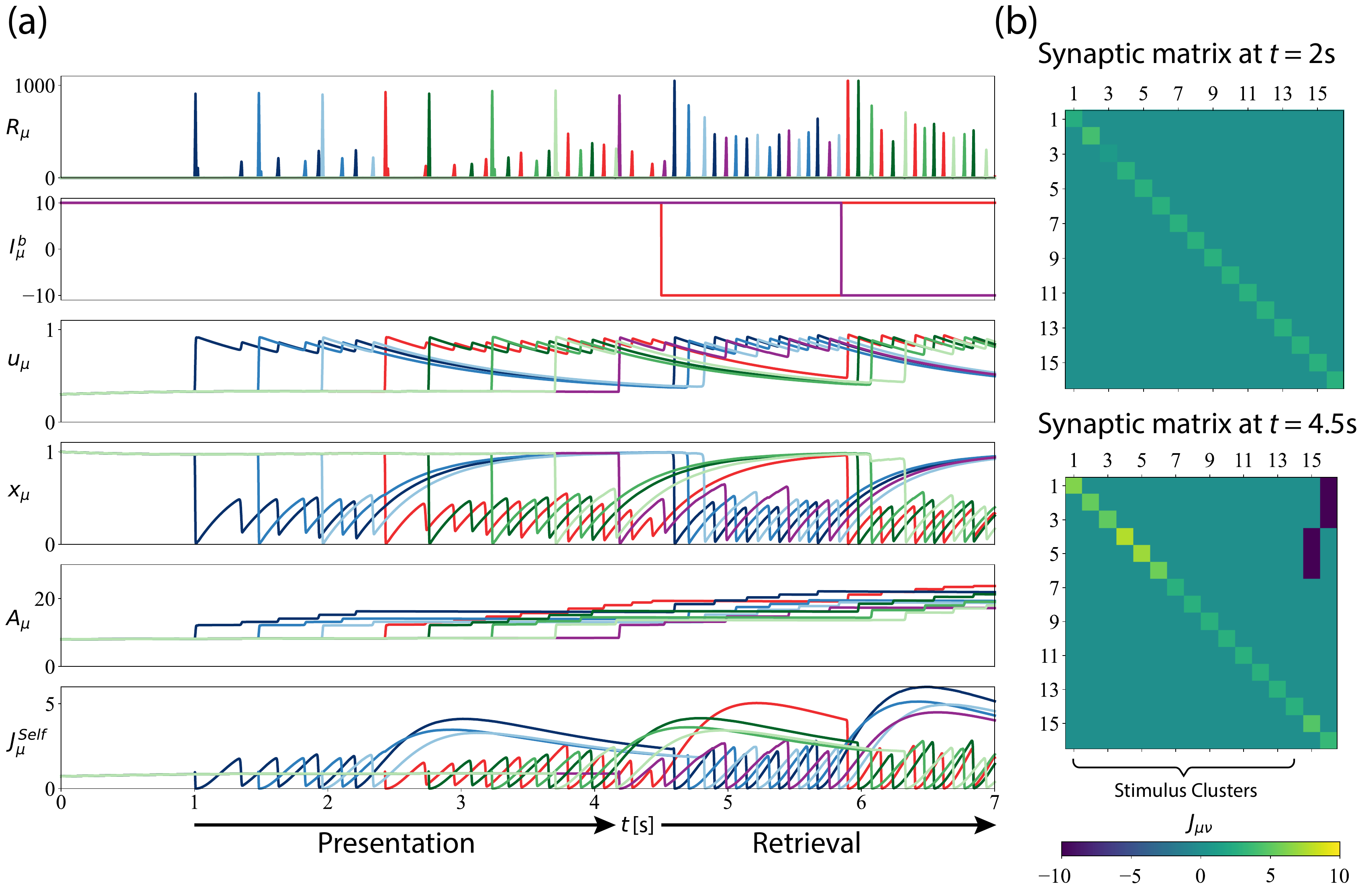}
    \caption{\textbf{Full activity trace of the bottom panel in Fig.~\ref{fig:schematics}(d).} \\
    \textbf{(a)} Activity traces of all variables. From top to bottom: firing rates $R_{\mu}$, background input currents $I^b_{\mu}$, release probability $u_\mu$, fraction of available neurotransmitters $x_\mu$, amplitude of the recurrent strength $A_\mu$, and effective recurrent self-connection strength $J^{\text{Self}}_\mu:= J_{\mu \mu} = u_\mu x_\mu A_\mu$. \\
    \textbf{(b)} Snapshots of the synaptic matrix $J_{\mu \nu}$ before and after chunking. Clusters 1-14 are stimulus clusters and 15-16 are chunking clusters. At $t=2\ \mathrm{s}$, only the first chunk (the blue colors) is presented, chunking clusters are not activated, and only the recurrent self-connections are nonzero in the synaptic matrix. At $t=4.5\ \mathrm{s}$, both chunks are formed by the chunking connections (dark blue colors in the top right corner).}
\label{appfig:fig1_extended}
\end{figure*}
\begin{figure*}[t!]
    \centering
    \includegraphics[width=1\textwidth]{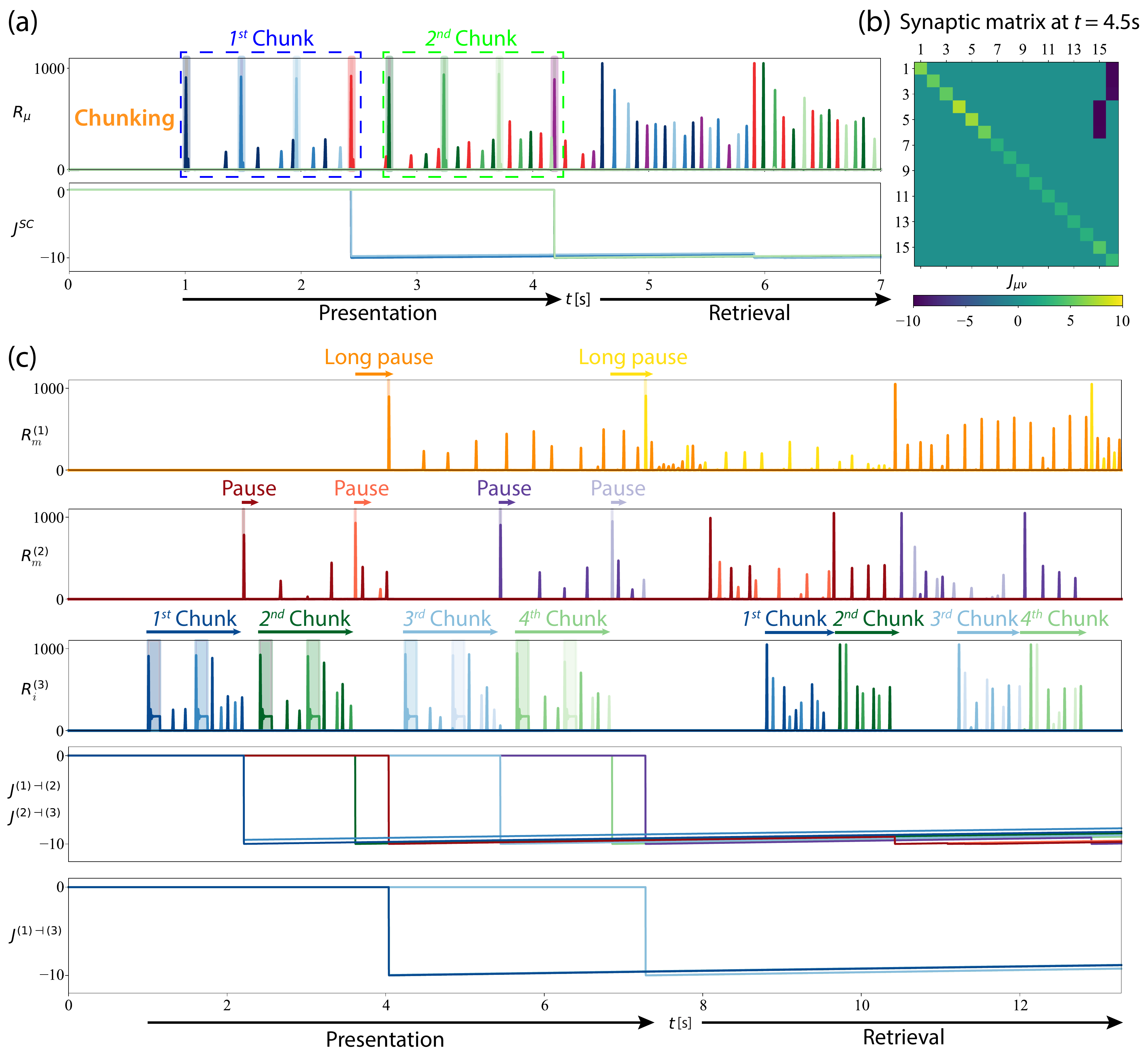}
        \caption{\textbf{Additional RNN simulations with delayed Hebbian plasticity.} \\
        \textbf{(a)} Approximating the chunking dynamics in Fig.~\ref{fig:schematics}(d) using Eq.~\eqref{eq:binding} instead of Eq.~\eqref{eq:inhibition}. Top: activity traces of the firing rates. Bottom: activity traces of the inhibitory connections from chunking clusters to stimulus clusters $J^{SC}$. \\
        \textbf{(b)} Snapshot of the synaptic matrix after chunking, resulting from the dynamics described in Eq.~\eqref{eq:binding}. \\
        \textbf{(c)} Approximating the chunking dynamics in Fig.~\ref{fig:hierarchy}(b) using Eq.~\eqref{eq:binding} instead of Eq.~\eqref{eq:inhibition}. Synaptic matrix components that correspond to the inhibition from level $k$ to $l$ are collectively denoted as $J^{(k) \dashv (l)}$. First three panels: firing rate activity traces of the clusters in Fig.~\ref{fig:hierarchy}(a). Fourth and fifth panels: inhibitory connections between adjacent levels $J^{(1)\dashv (2)}$ and $J^{(2)\dashv (3)}$, inhibitory connections between skip levels $J^{(1)\dashv (3)}$, resulting from the dynamics described in Eq.~\eqref{eq:binding}. }
\label{appfig:memory_trace}
\end{figure*}
\begin{figure*}[t!]
    \centering
    \includegraphics[width=1\textwidth]{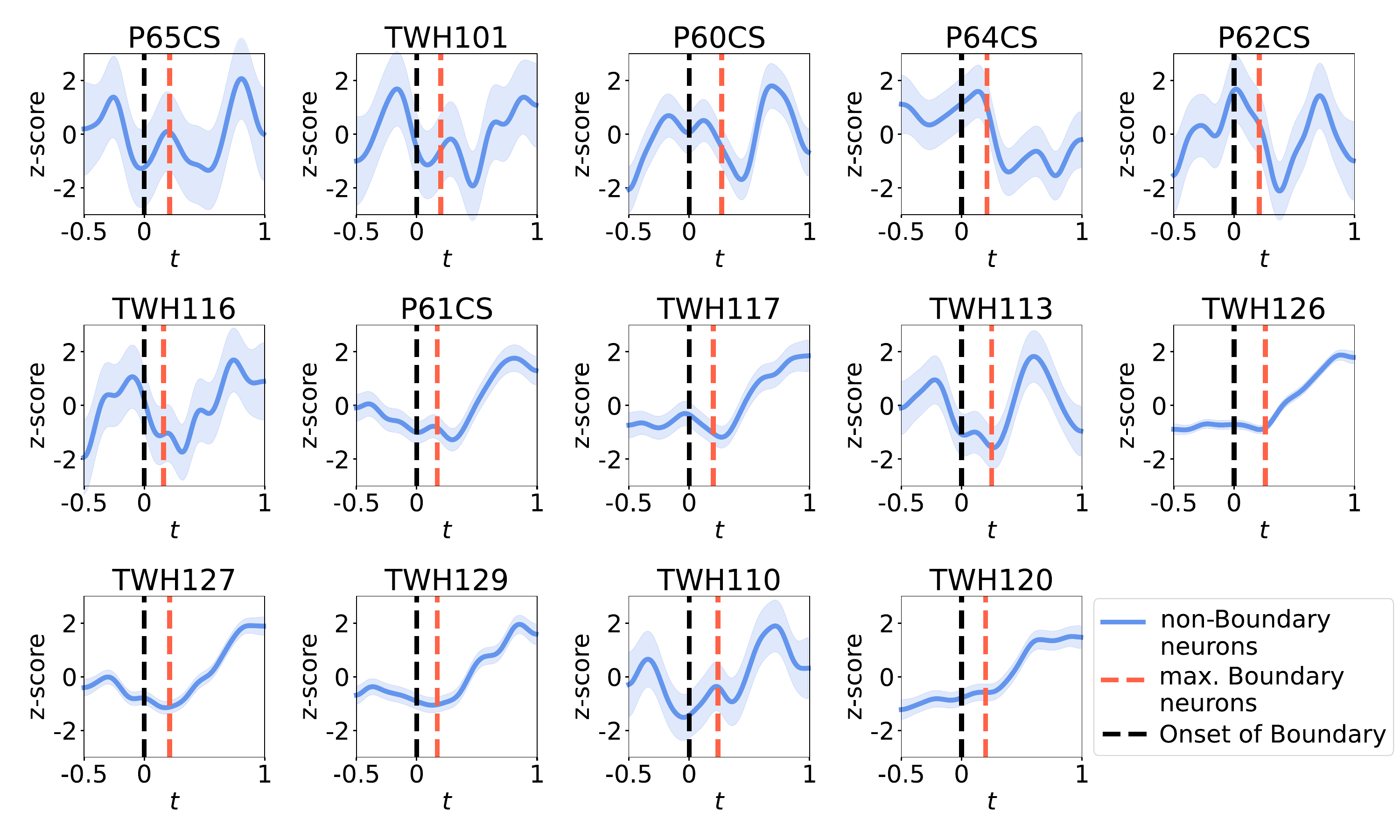}
        \caption{\textbf{Individual 2D plots of Fig.~\ref{fig:boundary_neurons}(b).} \\
        Individual subjects' z-score firing rates of the non-boundary neurons are shown in blue, with one standard deviation included as shades. Black dashed lines denote $t=0\ \mathrm{s}$ where the movie cut occurs. Red dashed lines denote the location of the maximum firing rate of the boundary neurons. Results are pooled from the raw firing rates of all non-boundary neurons from that subject. Subject IDs are presented according to the data in \cite{zheng2022neurons}. While some subjects do not exhibit the qualitative trend as predicted (e.g., the firing rate of subject P64CS does not have a ramp, and TWH120 does not have a dip), most of the subjects' firing rates follow the same qualitative trend as observed in the average plot in Fig.~\ref{fig:boundary_neurons}(a).}
\label{appfig:fig3_extended}
\end{figure*}

\clearpage

\pagebreak

\bibliography{ref.bib}

\end{document}